# Characterization of extrasolar terrestrial planets from diurnal photometric variability

E. B. Ford*, S. Seager† & E. L. Turner*

*Princeton University Observatory, Peyton Hall, Princeton, NJ 08540, USA
†Institute for Advanced Study, Einstein Drive, Princeton, NJ 08540, USA

**The detection of massive planets orbiting nearby stars has become almost routine[1,2], but current techniques are as yet unable to detect terrestrial planets with masses comparable to the Earth's. Future space-based observatories to detect Earth-like planets are being planned. Terrestrial planets orbiting in the habitable zones of stars—where planetary surface conditions are compatible with the presence of liquid water—are of enormous interest because they might have global environments similar to Earth's and even harbor life. The light scattered by such a planet will vary in intensity and colour as the planet rotates; the resulting light curve will contain information about the planet's properties. Here we report a model that predicts features that should be discernible in light curves obtained by low-precision photometry. For extrasolar planets similar to Earth we expect daily flux variations up to hundreds of percent, depending sensitively on ice and cloud cover. Qualitative changes in surface or climate generate significant changes in the predicted light curves. This work suggests that the meteorological variability and the rotation period of an Earth-like planet could be derived from photometric observations. Other properties such as the composition of the surface (e.g., ocean versus land fraction), climate indicators (for example ice and cloud cover), and perhaps even signatures of Earth-like plant life could be constrained or possibly, with further study, even uniquely determined.**

NASA and ESA are now considering two ambitious space missions—TPF[3] and Darwin[4] respectively—to detect and characterize terrestrial planets orbiting nearby Sun-like stars. Although very different designs are being considered[3,4,5,6], all have the goal of spectroscopic characterization of the atmospheric composition, and in particular the detection of gases that are important for or caused by life on Earth, including $O_2$, $O_3$, $CO_2$, $CH_4$ and $H_2O$[7]. A mission capable of measuring these spectral features would necessarily obtain sufficient signal-to-noise ratios to measure photometric variability. Many other important properties of an extrasolar planet could be derived from photometric measurements; this is the motivation for the investigation presented here. Photometric variability would be especially valuable for studying many planets quickly or a planet that is too dim for spectroscopic studies. Moreover, the photometric variability could be monitored concurrently with a spectroscopic investigation, as was done for the transiting extrasolar giant planet of HD209458[8].

We have developed a code which calculates the total light scattered by an extrasolar planet towards an observer. The code performs Monte Carlo integrations, with single scattering, over a spherical planet using a map which specifies the scattering surface type at each point on the sphere; the code also uses a set of wavelength-dependant bidirectional reflectance distribution functions that specify the probability that light incident from one direction will scatter into another direction for each type of scattering surface (refs 19-19; see also B. Rock and J. Salisbury http://speclib.jpl.nasa.gov/archive/jhu/becknic/vegetation/txt/decidous.txt). The observed flux also depends on the viewing geometry which is specified by the phase angle (the angle between the star, planet and observer), obliquity (the direction of the planet's rotation axis), orbital inclination relative to the line of sight, and time of day ($t$). Although the geography and climate of an Earth-like extrasolar planet are completely unknown, we can attempt to calculate what an optical or near-infrared light curve of Earth would look like if viewed from a very large distance. Simple variations about this basic reference model then allow us to explore reasonable possibilities for other Earth-like planets. We use a map of Earth from a one square degree satellite surface map which classifies each pixel as permanent ice, dirty/temporary ice, ocean, forest, brush, or desert[20]. We consider cloudy models separately, using the scattering properties of Earth clouds[21]. We focus our attention to quadrature (a phase angle of 90°) for which the planet-star separation is largest and the observational constraints thus least severe.

The diurnal light curve from our cloud-free Earth model has variation as high as 150% (Fig. 1). The significant intensity variation is due to the facts that surfaces have different albedos and a relatively small part of the visible hemisphere dominates the total flux from an unresolved planet. For example, at quadrature less than 10% of the surface often produces more than 50% of the total reflected light. In our cloud-free Earth model the variability is primarily due to land and ocean rotating in and out of view. The peak in the light curve at $t = 0.5$ day is caused by the high albedo of sand from the Sahara desert, while the dip at $t = 0.8$ day is caused by South America rotating into the location where there is usually specular reflection off of the ocean. Ice, sand, oceans, and vegetation can all produce significant features in the rotational light curve, but distinguish themselves by their colours (e.g. ice is very nearly grey, while sand has an albedo which rises by about 30% from 450 nm to 750 nm).

On Earth clouds are extremely important to the reflected light curve both because of their very high albedos and their short timescales of formation, motion and dissipation. Figure 2 shows the diurnal light curve of our Earth model with daily, seasonal, and annual average cloud coverage maps from the ISCCP database (ref. 22 and see http://isccp.giss.nasa.gov/). Here the intensity variation is due to the contrast between clouds and land or ocean. Clouds tend to raise the overall brightness and variability, reducing the fractional variation in the reflected light curve compared to the cloud-free case (Fig. 1). On Earth cloud patterns cause a variation of about 20% within a typical day. Cloud patterns can be coherent over several days (Fig. 2a), allowing the rotational period of the Earth to be measured from its light curve and thus the light curve to be averaged over many days. Light curves of Earth with seasonal and annual cloud coverage maps (Fig. 2b) demonstrate that average cloud patterns for Earth vary with seasons. These seasonal light curves reflect the fact that there are locations that are almost always cloudy (such as the Amazon Basin) and regions that are virtually cloud free (such as the Sahara desert) at different times of the year.

We could learn about Earth's unresolved rotationally modulated flux from a satellite far from Earth or by observing the reflection of the Earth from the dark side of the moon (Earthshine). Goode et al.[23] have measured Earthshine for about 200 days since 1998. Our model agrees with their measurements for both the mean reflectivity (0.086 model vs. 0.092 observed) and fractional variance (13% model vs. 15% observed) at a phase angle between 80° and 100°. Their observations are made from a single location, limiting their range of viewing geometries. Monitoring Earthshine from several different longitude locations will allow for more complete daily coverage.

While the Goode et al. Earthshine observations[23] affirm the accuracy of our Earth model, they cannot address the light curves of extrasolar planets which will not be identical to those of Earth. The rotational period, the continental fraction and arrangement could be completely different. In addition, the cloud fraction and spatial and temporal distribution depends on many variables such as the location of continents and obliquity. Since predicting cloud coverage is very difficult for global climate models, we use cloud-free models to consider plausible Earth-like planets by varying the surface map (Fig. 3). Figure 3 shows daily light curves for planets with several different surface maps which illustrate that models with qualitatively different surfaces produce distinctly different light curves. For example, an Earth-like planet whose land was covered by ice (Fig. 3a) or thick forests (Fig. 3b) would have much larger amplitude variations than the Earth's. Varying the fractional ocean coverage (Fig. 3c) affects both the normalization and the variability of the light curve. Wheras Figure 3c and d demonstrates that an extrasolar planet's diurnal light curve may contain information about the planet's surface, the full inverse problem of obtaining a unique determination of surface features from a light curve may be intractable and will certainly require much more investigation. For example, changing the obliquity of our cloudless Earth model also causes significant changes in the light curve (Fig. 3d). Nevertheless, a planet's light curve will clearly place constraints on its surface properties and climate in that many possible models would be ruled out by any specific set of photometric data.

In addition to studying 'geological' surface features we might learn about biological features encoded on the planet's surface. On Earth vegetation has a dramatic sudden rise in albedo by almost an order of magnitude around 750 nm, known as the red edge. Vegetation has evolved this strong reflection as a cooling mechanism to prevent overheating which would cause chlorophyll to degrade. Although the albedo of sand also increases towards the infrared, the red edge vegetation signature is more rapid and may be detectable from the unresolved Earth. (See also the Galileo observations of part of Earth[24].) We cannot necessarily expect to find Earth-like vegetation on extrasolar terrestrial planets, but photometric measurements in different colours may be able to detect a unique signature, different from any known surface features or atmospheric constituents on Earth or other solar system planets.

We expect the diurnal rotational variation of an Earth-like planet to be lower in the mid-infrared (flux variation of a few percent) than in the optical because the surface temperature does not vary as much as surface albedo across the Earth. For a planet with nonzero obliquity the mid-infrared seasonal flux variation should be larger than mid-infrared rotational variation because of seasonal temperature variation. In addition, optical multicolour photometry and polarization as a function of phase angle (on an orbital timescale) could help constrain the type of atmospheric scatterers and the particle size distribution or detect a large smooth specular surface such as an ice- or an ocean-covered planet. Alternatively, the variation on an orbital timescale may be overwhelmed by variations indicating strong seasonal changes in the atmosphere.

Theoretical light curves of the unresolved Earth have 10-20% variation. A TPF that can measure 5% optical variation could detect weather, the rotational period, and seasonal changes in the cloud pattern of a planet like our Earth. If the surface were to contribute most of the scattered light rather than the clouds, then planets with different surface features would show very different diurnal light curves. This is in dramatic contrast to planets like Venus which would show almost no diurnal variability. Thus, we expect the diurnal light curve of an extrasolar Earth-like planet to contain detectable features encoding information about its physical and perhaps even biological properties.

We thank the members of the Ball Aerospace Terrestrial Planet Finder Team, B. Soden, A. Broccoli and G. Williams for useful discussions, and B. Soden for assistance with the ISCCP database. We wish to thank W. B. Rossow, the Goddard Institute for Space Studies, the Goddard Space Flight Center, and NASA for the production and distribution of this data set. E.B.F. would like to acknowledge support under a National Science Foundation Graduate Fellowship and S.S. is supported by the W. M. Keck Foundation.

**Correspondence and requests for materials should be addressed to E.B.F. (e-mail: eford@astro.princeton.edu).**

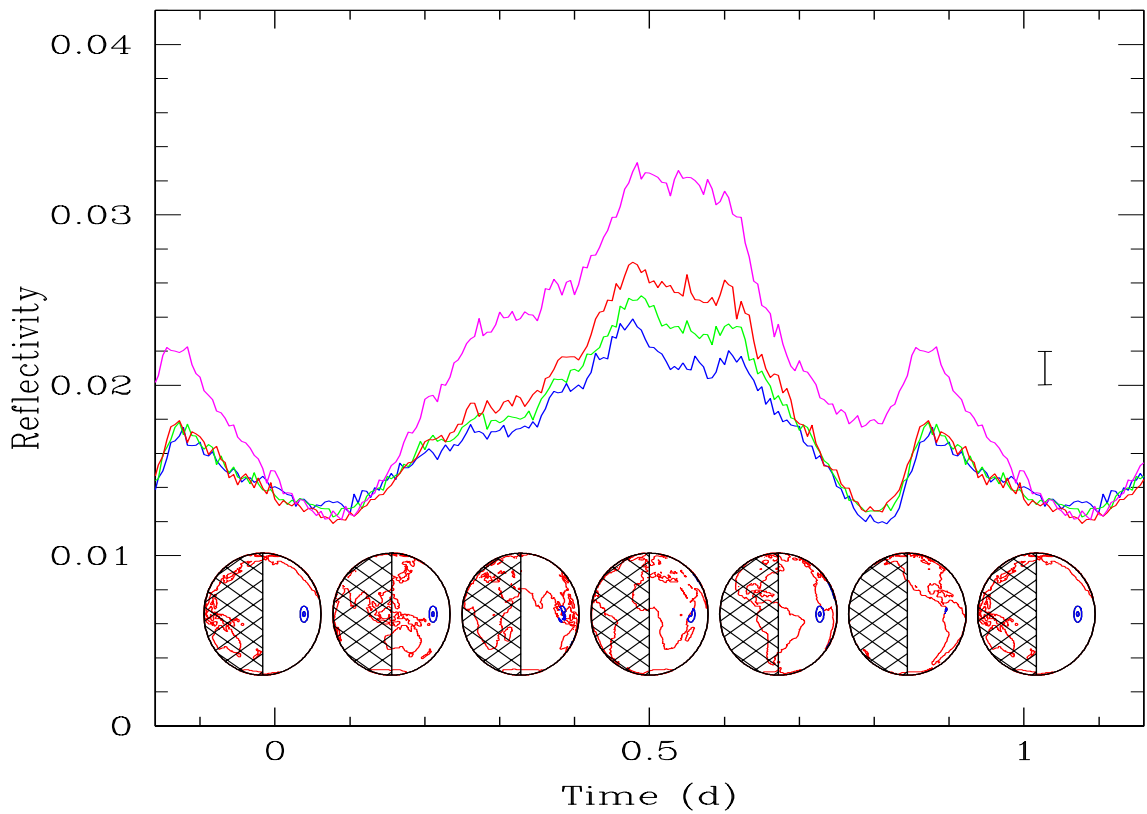

Figure 1. Rotational light curve for a cloud-free Earth model. The light curve is sampled at four-minute intervals and the Poisson noise is due to Monte Carlo statistics in our calculations. The pink, red, green, and blue curves correspond to wavelengths of 750, 650, 550 and 450 nm, and their differences reflect the wavelength-dependent albedo of different surface components. The difference in magnitude of the star and planet is $\Delta m = 22.60 + 5 \log(r) - 5 \log R_p - 2.5 \log \Re$, where $r$ is the star-planet separation in astronomical units, $R_p$ is the radius of the planet in Earth radii, and $\Re$ is the reflectivity normalized to a Lambert disk at a phase angle of 0° which is plotted on the y-axis. Note that a different phase angle affects $\Re$ due to a larger or smaller fraction of the disk being illuminated. The images below the light curve show the viewing geometry (the cross-hatched region is not illuminated), a map of the Earth (red), and the region of specular reflection from the ocean (blue). At $t = 0.5$ day, the Sahara desert is in view and causes a large peak in the light curve due to the reflectivity of sand which is especially high in the near-infrared (pink curve). The error bar in this and subsequent figures shows an estimate of the photometric accuracy of a TPF mission.

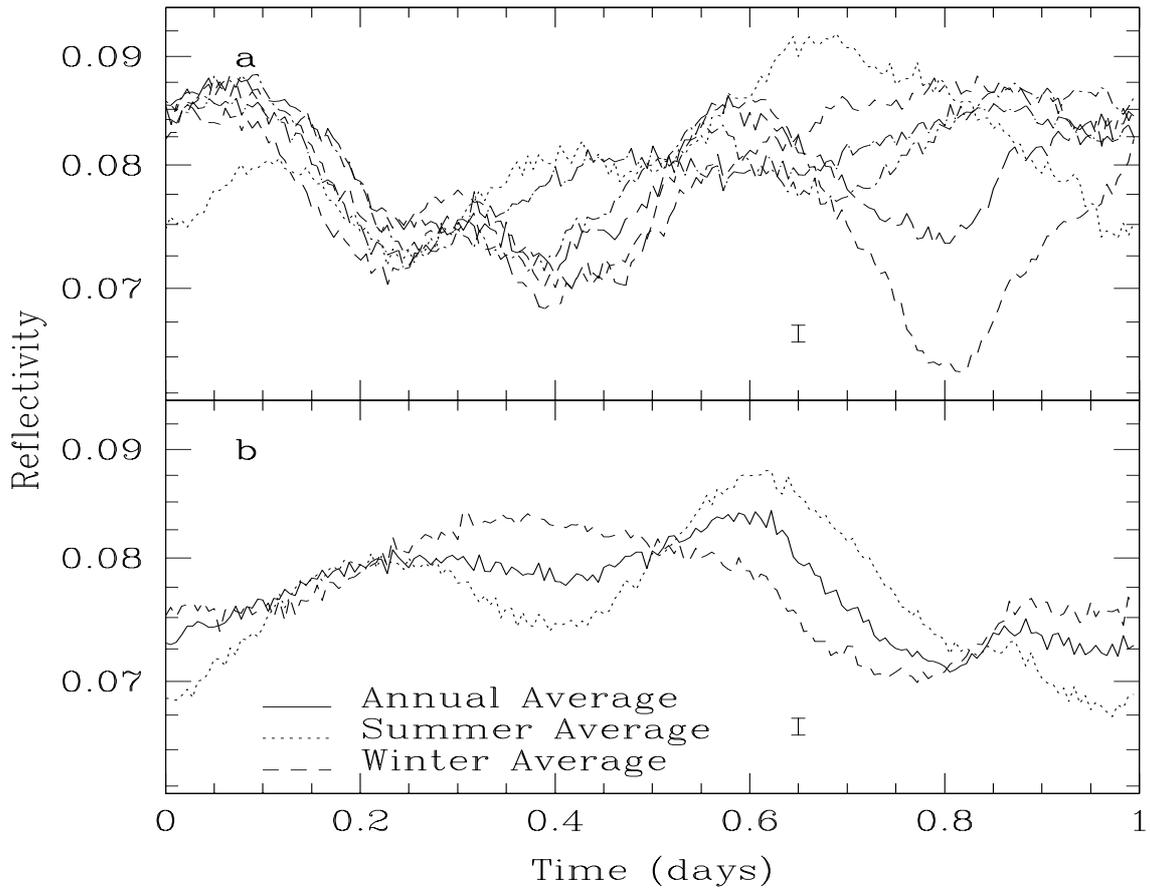

Figure 2. Rotational light curves for model Earth with clouds. Model daily light curves at 550 nm for our Earth model with clouds, as viewed from quadrature. Panel a shows theoretical light curves using cloud cover data from satellite measurements taken on six consecutive days (13-18 April 1986). Panel b shows theoretical light curves for Earth using seasonal (dotted and dashed lines) and annual (solid line) average cloud cover (averaged over 8 years). Using actual cloud data allows us to accurately model the Earth, but is not applicable to extrasolar terrestrial planets.

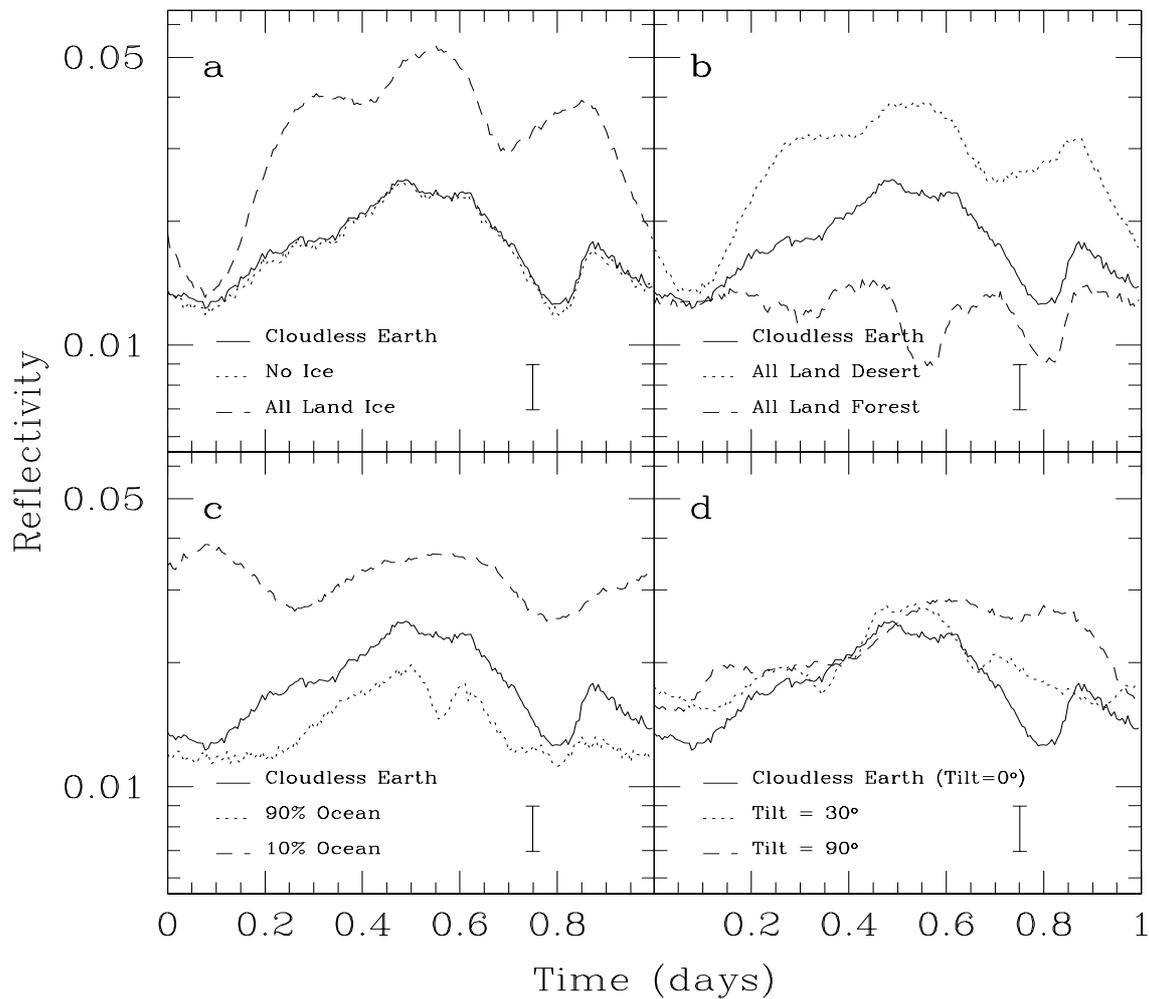

Figure 3. Rotational light curves for cloudless Earth-like models. Model daily light curves at 550 nm for an Earth-like planet, as viewed from quadrature. The light curves shown in panels a-c are generated using variations on our cloud-free Earth model which is presented as the solid black line in each panel for reference (corresponding to the green curve in Figure 1). These models do not include clouds because the cloud coverage would differ for each Earth-variant in a way not yet known. Panel a, Light curves for Earth-like planets for which all the land is covered with ice (dashed line) and none of the land is covered with ice (dotted line). Panel b, Light curves for Earth-like planets for which all the land is covered with thick forests (dashed line) and all of the land is covered with desert (dotted line). Panel c, Light curves for Earth-like planets for which the land and oceans have been recursively expanded and contracted along their perimeters to achieve the desired fractional ocean coverage. Panel d, Light curves for our cloudless Earth model for different obliquities. The north pole has been tilted towards the Sun by 0° (solid line), 30° (dotted line), and 90° (dashed line).